\documentclass[12pt]{article}
\usepackage{epsf}
\author{ A.\,Samsonov \\\\
\it {\small{Institute of Theoretical and Experimental Physics,}} \\
\it {\small{Bol'shaya Cheremushkinskaya, 25, Moscow, 117259, Russia}}}
\bf
\title{Gluon condensate in charmonium sum rules \\for
the axial-vector current}

\begin{document}
\date{}
\maketitle
\newcommand{\qq}{\langle\overline{q}q\rangle^2}

\def\la{\mathrel{\mathpalette\fun <}}
\def\ga{\mathrel{\mathpalette\fun >}}
\def\fun#1#2{\lower3.6pt\vbox{\baselineskip0pt\lineskip.9pt
\ialign{$\mathsurround=0pt#1\hfil##\hfil$\crcr#2\crcr\sim\crcr}}}

$$\centerline{\hbox{Abstract}}$$ 
{\small{
\indent The charmonium sum
rules for the axial-vector current are considered. The three-loop
perturbative corrections, operators up to 8 dimension and
$\alpha_s$-corrections to the lowest dimension operator are accounted.
The contribution of the operators of 6 and 8 dimensions is
computed in the framework of factorization hypothesis and
instanton model.
For the value of gluon condensate the following result is
 obtained:
$\langle{\alpha_s\over\pi}G^2\rangle=(0.005+0.001-0.004)\,{\rm
GeV}^4$ (for the charm quark mass ${\bar m}=1.275\pm 0.015 \,{\rm
GeV}$). }}
\\
\newpage
$\,$\\
\section{Introduction}
$\,$ \indent It is well known that the nonperturbative phenomena in QCD 
are of great importance. Indeed, quark and gluon condensates determine
the properties of hadrons and their interactions to a considerable
extent. This was stressed by Shifman, Vainshtein and Zakharov in
\cite{svz}, where the technique of QCD sum rules was proposed. \\
\indent In particular, authors of \cite{svz} pointed out to
the significance of the gluon condensate in nonperturbative QCD.
First of all, gluon condensate has the lowest dimension among
chirality conserving  condensates and that is why it plays dominant 
role in the
sum rules, corresponding to the processes without chirality
violating. Moreover, the gluon condensate is directly
related to the density of vacuum energy. \\
\indent The numerical value of
the gluon condensate was found in \cite{svz} from the analysis of
charmonium sum rules. It was 
$$\langle{\alpha_s\over\pi}G_{\mu\nu}^a
G_{\mu\nu}^a\rangle=0.012\, {\rm GeV}^4.$$ 
In this
estimation the value $\Lambda_{QCD}=100 {\rm MeV}$ was used and
$\alpha_s$-corrections of the first order were taken into account.
\\ \indent However, at the present moment $\Lambda_{QCD}$ is known
to be sufficiently larger, \\ $\Lambda_{QCD}\approx 250\,{\rm
MeV}$, and, furthermore, $\alpha_s$-corrections of the second
order to the polarization operator are available. In addition, 
authors of \cite{svz} worked at $Q^2=0$, where the higher order
terms in the operator expansion series are significant
\cite{nik-rad-1}. These facts are taken into account in the recent
paper \cite{iof-zya}, where the charmonium sum rules for the
vector current are considered over again with the purpose to
obtain the value of $\langle{\alpha_s\over\pi}G_{\mu\nu}^a
G_{\mu\nu}^a\rangle$. (Of course, there were many other papers,
devoted to the calculation of the gluon condensate, the short list
of them can be found in \cite{iof-zya}, 
for review see \cite{iof-condens}). \\ 
\indent As the result,
 authors of \cite{iof-zya} obtained
$\langle{\alpha_s\over\pi}G_{\mu\nu}^a G_{\mu\nu}^a\rangle=
 0.009\pm 0.007\,{\rm GeV}^4$ and ${\bar m}=1.275\pm 0.015\, {\rm GeV}$
for the $c$-quark mass in $\overline{\rm MS}$ scheme. 
Thus one can see that the accuracy of the
quark mass value is high, whereas the error in the value of the
gluon condensate is comparable with its magnitude. \\ 
\indent In
the paper \cite{zya} the charmonium sum rules for the pseudoscalar
current were analysed, and for the gluon condensate value
the following restriction was obtained: \\
$\langle{\alpha_s\over\pi}G_{\mu\nu}^a G_{\mu\nu}^a\rangle \la
 0.008\, {\rm GeV}^4$.
\\
\indent In the present paper we try to obtain the gluon condensate by
considering other independent channel in charmonium, namely, we
analyze the axial-vector current. \\
\indent It should be mentioned that the values of gluon condensate and
$c$-quark mass are interdependent in the charmonium sum rule, i.e. the
variation of one of them results in the changing of another. That is why
the determination of independent restriction to each of them is a
special problem. In the present paper we use the quark mass
value, obtained in
\cite{iof-zya}, as the input parameter and devote our attention to the gluon
condensate. \\
\section{Correlator of the axial-vector currents}
\indent We consider the correlator of the charmed quark axial-vector currents:
$$i\int d^4x\,e^{ipx}\langle T(j_\mu(x)j^+_\nu(0)\rangle=(q_\mu q_\nu
-g_{\mu\nu}q^2)\Pi(q^2)+q_\mu q_\nu \Pi_L(q^2)\,.\eqno(1)$$
Here $j_\mu=\overline{c}\gamma_\mu \gamma_5 c$, $\Pi_L$ is the longitudinal
part of the polarization operator. In what follows, only transverse part
of the polarization operator $\Pi(q^2)$ is considered. \\
\indent $\Pi(q^2)$ (1) can be
expressed through its imaginary
part with the help of dispersion relation:
$$\Pi(q^2)={q^2\over{4\pi^2}}\int\limits_{4m^2}^\infty
{R(s)ds\over{s(s-q^2)}}\,,$$
where
$$R(s)=4\pi {\rm Im}\Pi(s+i0)\,,\eqno(2)$$
$m$ is the pole mass of $c$-quark. \\
\indent The dispersion relation
can be saturated by the contributions of physical states. In the
axial-vector channel the only resonance with the mass
$m_\chi=3510.51\pm 0.12 MeV$ is known \cite{pdg}. We use the
simplest model of the spectrum, containing this resonance and
continuum. As usual, in order to suppress the continuum
contribution one  should consider the derivatives of the
polarization operator in Euclidean region $Q^2=-q^2>0$:
$$M_n(Q^2)={4\pi^2\over{n!}}\Big(-{d\over{dQ^2}}\Big)^n \Pi(Q^2)=
\int\limits_{4m^2}^\infty{R(s)ds\over{(s+Q^2)^{n+1}}}\,.\eqno(3)$$
Usually the quantities $M_n$ are referred to as moments. \\ 
\indent
From the phenomenological point of view
$$M_n(Q^2)=4\pi^2\sum\limits_\chi{|\langle
 0|j_\mu|0\rangle|^2\over{(m_\chi^2+Q^2)^{n+1}}}\,,$$ and
 in the ratio of two successive moments the contribution
of the lightest state dominates at large $n$:
$${M_{n-1}(Q^2)\over{M_n(Q^2)}}=m_\chi^2+Q^2+\delta_c\,,$$ where
$\delta_c$ is the continuum contribution. \\
  \indent The QCD part of the polarization
operator consists of the
 perturbative and nonperturbative terms:
$$\Pi(q^2)=\Pi^{Pert}(q^2)+\Pi^{OPE}(q^2)\,.$$
\indent The first one is determined by its imaginary
part (2). $R(s)$ can be expressed as the series in the coupling constant
$\alpha_s$:
$$R(s)=\sum\limits_{k=0,1,...}R^{(k)}(s,\mu^2)a^k(\mu^2)\,.\eqno(4)$$
Hereafter we denote $a(\mu^2)=\alpha_s(\mu^2)/\pi$. We
will take first three terms in this series. \\ 
\indent $R(s)$ is
the physical quantity and does not depend on the scale $\mu^2$,
but each term in (4) can depend. First two terms of (4) are known
analytically. They do not contain scale dependence. One can find
them in \cite{hoa-teub}:
$$\displaylines{R^{(0)}=v^3\,,\hfill(5)}$$
$$\displaylines{R^{(1)}={4\over 3}\Bigg(v^2\Big( (1+v^2)(2\ln(1-p)\ln p
+\ln(1+p)\ln p +2{\rm Li_2}(p)+{\rm Li_2}(p^2))-\hfill}$$
$$\displaylines{-2v(2\ln(1-p)+\ln(1+p))\Big)-{1-v\over{32}}(21+21v+80v^2
-16v^3+3v^4+3v^5)\ln p+\hfill(6)}$$
$$\displaylines{\hfill+{3v\over{16}}(-7+10v^2+v^4)\Bigg)\,.~~~~~~}$$
In these expressions $v$ is the quark velocity, $v=\sqrt{1-4m^2/s}$, 
$p=(1-v)/(1+v)$. ${\rm Li_k}$
is the polylogarithm function:
$${\rm Li_k}(v)=\sum\limits_{n=1}^\infty{v^n\over{n^k}}\,.$$
\indent The term $R^{(2)}$ is represented usually as the sum of five gauge
invariant parts:
$$R^{(2)}=C_F^2R^{(2)}_A+C_A C_FR^{(2)}_{NA}+C_F Tn_l R^{(2)}_{l}+
C_F T R^{(2)}_{F}+C_F T R^{(2)}_{S}\,.\eqno(7)$$ Here
$C_A=3,\,\,C_F=4/3,\,\,T=1/2$ are group constants and $n_l=3$ is the
number of light quarks. \\ 
\indent The term $R^{(2)}_{l}$
corresponds to the so-called double-bubble diagram, the diagram
with two quark loops, in external loop there are massive quarks,
whereas in internal one -- massless quarks. $R^{(2)}_{l}$ has the
following form \cite{hoa-teub}:
$$R^{(2)}_{l}=\bigg(-{1\over 4}\ln{\mu^2\over{4s}}-{5\over{12}}\bigg)R^{(1)}
+\delta^{(2)}_A\,,\eqno(8)$$
where $\delta^{(2)}_A$ is given by equation (79) in \cite{hoa-teub}. \\
\indent $R^{(2)}_{F}$ appears from the double-bubble diagram with
massive quarks in both loops. In the domain $s<16m^2$ only the virtual
massive quarks contribute, and $R^{(2)}_{F}$ has the form \cite{hoa-teub}:
$$R^{(2)}_{F}=2v^3{\rm Re} F^{(2)}_{3,Q} -{1\over
 4}\ln{\mu^2\over{m^2}}R^{(1)}\,.\eqno(9)$$ The expression for
$F^{(2)}_{3,Q}$ can be found in appendix B of \cite{hoa-teub},
equation(166). \\ 
\indent For $s>16m^2$ the contribution of the
real quarks appears. It is given by double integral, which can not
be taken analytically (equation (71) in \cite{hoa-teub}). The
numerical calculation shows that it is small, nevertheless, we
will take it into account. \\ 
\indent The functions $R^{(2)}_A$
and $R^{(2)}_{NA}$ appear from diagrams with single quark loop
and a number of gluon lines, they contain abelian and nonabelian
exchanges correspondingly. These functions are not known
analytically, consequently, one has to use some approximations. \\
\indent Our approximation expressions are based 
on the first eight moments at
$Q^2=0$, which are known analytically \cite{ch-kuhn-st}. The
obtained formulas, being substituted into (3), have to
reproduce  these eight moments with high accuracy. In order to
construct such approximations we perform the following steps. \\
\indent Let us consider the series
$$\Pi^{(2)}_{NA}={3\over{16\pi^2}}\sum\limits_{k=1,2,...}C^{(2)}_{NA,k}
\,z^k\,,$$ where
$z=q^2/(4m^2)$,
$\Pi^{Pert}=\sum\limits_k\Pi^{(k)}a^k$ and  \\
$\Pi^{(2)}=C_F^2\Pi^{(2)}_A+C_A C_F\Pi^{(2)}_{NA}+C_F Tn_l
\Pi^{(2)}_{l}+ C_F T \Pi^{(2)}_{F}+C_F T \Pi^{(2)}_{S}$ similar to (7). The
coefficients $C^{(2)}_{NA,1},...,C^{(2)}_{NA,8}$ are known
analytically. First of all, we reexpand this series in terms of
variable $\omega$,
$$\omega={1-\sqrt{1-z}\over{1+\sqrt{1-z}}}\,.$$
Thus we map the complex $q^2$-plane to the unit circle. Then we
construct the Pade approximation, which usually gives a better
accuracy than Tailor series. It has the following form:
$$\Pi(\omega)={a_0+a_1\omega+...+a_i\omega^i\over{1+b_1\omega+...+
b_j\omega^j}}\,.$$ Since we have 8 moments in hand, we can
construct Pade approximation with 8 parameters $a_i$ and $b_j$. The
best results turns out to give the following approximation:
$$\displaylines{\Pi^{(2)}_{NA}(\omega)={3\over{16\pi^2}}\times\hfill}$$
$$\displaylines{\hfill\times
{10.8547\,\omega+9.43221\,\omega^2-8.76722\,\omega^3-
 1.74256\,\omega^4+0.853743\,\omega^5-0.257734\,\omega^6\over{
 1+0.373686\,\omega-0.439076\,\omega^2}}\,.~~(10)}$$
\indent In the similar way we obtain for the abelian part:
$$\displaylines{\Pi^{(2)}_{A}(\omega)={3\over{16\pi^2}}\times\hfill}$$
$$\displaylines{\hfill\times
{9.25606\,\omega-288.334\,\omega^2-394.513\,\omega^3-
 69.8439\,\omega^4+19.9673\,\omega^5-0.936404\,\omega^6
\over{1-32.0000\,\omega-15.2803\,\omega^2}}\,.~~(11)}$$
\indent The last term in (7) is generated by the diagram with two triangle
quark loops (singlet part). We need the first moments in the expansion
$$\Pi^{(2)}_{S}={3\over{16\pi^2}}\sum\limits_{k=1,2,...}C^{(2)}_{S,k}
\,z^k\,.\eqno(12)$$ 
Such diagrams for 
different currents were discussed in paper \cite{ch-harl-st}. 
However, to cancel the axial anomaly, in the
axial-vector case the current $\tilde j_\mu=
\overline{c}\gamma_\mu \gamma_5 c-\overline{s}\gamma_\mu \gamma_5
s$ was considered. As the result, in the calculated factors
${\tilde C}^{(2)}_{S,k}$ ($k=1,...,7$) the logarithms
$\ln(q^2/m^2)$ appear due to massless cut  (see
equation (8) in \cite{ch-harl-st}). For our purposes we
need to subtract the contribution of the massless $s$-quark in
${\tilde C}^{(2)}_{S,k}$. The expression for the imaginary part
$R^{(2)}_{Ss}$ (the contribution of the massless cuts) 
 can be found in \cite{kniel}. But the
corresponding integral can not be taken analytically. The purely
numerical integration is also impossible because of the presence
of divergent logarithms $\ln(q^2/m^2)$ (which cancel out after
subtracting from ${\tilde C}^{(2)}_{S,k}$ ). The suitable
technique was suggested in \cite{ch-harl-st}. $R^{(2)}_{Ss}$ is
represented as the sum $$R^{(2)}_{Ss}={9\over
 2}\ln{s\over{m^2}}+R^{\prime(2)}_{Ss}\,.$$ Then the integral over
$R^{\prime(2)}_{Ss}$ is divided into three parts:
$$\int\limits_0^\infty {R^{\prime(2)}_{Ss}(r)\,dr\over{r-z}}=
\int\limits_0^\epsilon {R^{\prime(2)}_{Ss}(r)\,dr\over{r-z}}+
\int\limits_\epsilon^1 {R^{\prime(2)}_{Ss}(r)\,dr\over{r-z}}+
\int\limits_1^\infty {R^{\prime(2)}_{Ss}(r)\,dr\over{r-z}}\,.\eqno(13)$$
In the first integral on the right hand side we expand
$\tilde{R}^{\prime(2)}_{Ss}$ near $r=0$ and take about 50 terms to obtain a stable
sum of two first integrals in the range $\epsilon=0.65...0.75$
with the high accuracy. The third term on the right hand side of (13)
can be calculated purely numerically. Thus we obtain the coefficients in (12):\\
$C^{(2)}_{S,1}=-0.20665154$,\\
$C^{(2)}_{S,2}=-0.063212891$,\\
$C^{(2)}_{S,3}=-0.024202688$,\\
$C^{(2)}_{S,4}=-0.01062245$,\\
$C^{(2)}_{S,5}=-0.0050245436$,\\
$C^{(2)}_{S,6}=-0.002441955$,\\
$C^{(2)}_{S,7}=-0.0011555635$.\\
\indent On the basis of these
numbers one can obtain the following approximation:
$$\displaylines{\Pi^{(2)}_{S}(\omega)={3\over{16\pi^2}}\times\hfill}$$
$$\displaylines{\hfill\times{-0.826606\,\omega+0.179208\,\omega^2
+0.5387\,\omega^3-0.350049\,\omega^4+0.0425326\,\omega^5
\over{1+0.559635\,\omega-0.196815\,\omega^2}}\,.~~(14)}$$
\indent It should be noted that
the purely gluonic cut in this diagram is zero
according to the Landau-Yang theorem \cite{land-yang}. \\
\indent In the last step we take the imaginary part
$R_\delta^{(2)}$, $\delta=NA,\,A,\,S$ of the (10),(11),(14) ($\mu^2=m^2$):
$$R_\delta^{(2)}=4\pi\,{\rm Im}\Pi_\delta^{(2)}(\omega)\,,\eqno(15)$$
$$\omega={1+i\sqrt{z-1}\over{1-i\sqrt{z-1}}}\,.$$
$\,$\\
\indent The nonperturbative part of the polarization operator was calculated
up to the operators of dimension 8 in \cite{nik-rad}. For the correlator
of heavy quarks there are: \\
the only operator of dimension 4,
$$O_2=\langle g^2G_{\mu\nu}^aG_{\mu\nu}^a\rangle\,,$$
two operators of dimension 6,
$$O_3^1=\langle g^3f^{abc}G_{\mu\nu}^aG_{\nu\eta}^b
G_{\eta\mu}^c\rangle\,,\;\;\;\;O_3^2=\langle g^4j_\mu^aj_\mu^a\rangle\,,$$
and seven operators of dimension 8:
$$O_4^1=\langle(g^2d^{abc}G_{\mu\nu}^b G_{\eta\lambda}^c)^2 +{2\over 3}
(g^2 G_{\mu\nu}^a G_{\eta\lambda}^a)^2\rangle\,,\;\;\;\;
O_4^2=\langle(g^2f^{abc}G_{\mu\nu}^b G_{\eta\lambda}^c)^2\rangle\,,$$
$$O_4^3=\langle(g^2d^{abc}G_{\mu\eta}^b G_{\eta\nu}^c)^2 +{2\over 3}
(g^2 G_{\mu\eta}^a G_{\eta\nu}^a)^2\rangle\,,\;\;\;\;
O_4^4=\langle(g^2f^{abc}G_{\mu\eta}^b G_{\eta\nu}^c)^2\rangle\,,$$
$$O_4^5=\langle(g^5f^{abc}G_{\mu\nu}^a j_\mu^b j_\nu^c\rangle\,,\;\;
O_4^6=\langle(g^3f^{abc}G_{\mu\nu}^a G_{\nu\eta}^b
G_{\eta\mu;\lambda\lambda}^c\rangle\,,\;\;
O_4^7=\langle g^4j_\mu^aj_{\mu;\eta\eta}^a\rangle\,.$$
Here $j_\mu^a$ is the color current of the light quarks,
$gj_\mu^a=G_{\mu\nu;\nu}^a={g\over 2}\sum\limits_{q=u,d,s}
\overline{q}\gamma_\mu\lambda^a q$.
The operator product expansion part $\Pi^{OPE}$ of the polarization operator
 has the form \cite{nik-rad} ($y=Q^2/(4m^2)$,  
${}_2F_1(a,b,c,z)$ is hypergeometric function):
$$\displaylines{
\Pi^{OPE}(y)=
{1\over{4\pi^2}}{O_2\over{(4m^2)^2}}\Big({1\over{3y}}-{2\over 5}\,
{}_2F_1(1,4,7/2,-y)\Big)+\hfill}$$
$$\displaylines{~~~~~~~~~~~~~~~~~~~~~+{1\over{27\pi^2}}\sum\limits_{j=1}^2
{O_3^j\over{(4m^2)^3}}\Big({c_{3,0}^j\over y}+\sum\limits_{i=1}^5
c_{3,i}^j\,{}_2F_1(1,2+i,9/2,-y)\Big)+\hfill(16)}$$
$$\displaylines{\hfill+{1\over{27\pi^2}}\sum\limits_{j=1}^7
{O_4^j\over{(4m^2)^4}}\Big({c_{4,0}^j\over y}+\sum\limits_{i=1}^7
c_{4,i}^j\,{}_2F_1(1,3+i,11/2,-y)\Big)\,.}$$
In this equation
$$c_{3,0}^1=-3/5,\;\;c_{3,1}^1=0,\;\;c_{3,2}^1=0,\;\;
c_{3,3}^1=24/7,\;\;c_{3,4}^1=-116/7,\;\;c_{3,5}^1=552/35,$$
$$c_{3,0}^2=36/5,\;\;c_{3,1}^2=8/105,\;\;c_{3,2}^2=8/35,\;\;
c_{3,3}^2=144/35,\;\;c_{3,4}^2=32/21,\;\;c_{3,5}^2=-736/35,$$
$c_{4,i}^j=$
{\small{
$$\begin{array}{|c|cccccccc|}\hline
j\backslash i & 0 & 1 & 2 & 3 & 4 & 5 & 6 & 7 \\ \hline
 1 & 0 & 2/315 & -16/63 & 92/21 & 544/63 & -562/9 & 160/3 & 0 \\
 2 & 9/35 & 1/105 & 32/63 & -118/9 & 200/3 & -5639/45 & 30448/315 &
-180/7 \\
 3 & -18/5 & -8/315 & 64/63 & -752/21 & 9920/63 & -10036/45 &
 320/3 & 0 \\
 4 &  -144/35 & -2/315 & -32/45 & -188/63 & 7376/63 & -10586/45 &
-26464/315 & 1656/7 \\
 5 & 612/35 & -4/105 & -64/63 & 184/7 & -4000/21 & 20524/45 &
-1984/7 & -432/7 \\
 6 & 72/35 & 0 & -128/315 & 944/63 & -432/7 & -1544/45 & 98944/315 &
-1728/7 \\
 7 & -324/35 & 4/315 & 0 & 40/9 & -1696/63 & 292/3 & -14528/63 &
 1296/7 \\
\hline
\end{array}.$$
}}
\indent The $\alpha_s$-correction 
of the first order to the gluon condensate was
obtained in \cite{broad}. The corresponding
term in the polarization operator has the following form:
$$\Pi^{O2(1)}(Q^2)={1\over{4\pi^2}}{O_2\over{(4m^2)^2}}a(m^2)
{P^A(-y)\over{8y^2(1+y)}}\,,\eqno(17)$$
where $P^A(z)$ is given in equation (8) of \cite{broad}.
\section{Moments and mass redefinition}
$\,$
\indent In our approach the moments
contain the perturbative part up to $\alpha_s^2$
terms, the operator product expansion part with the operators of 4, 6 and 8
dimensions and perturbative correction of  the first order to the
lowest dimension operator:
$$M_n(Q^2)=\sum\limits_{k=0}^2 M_n^{(k)}(Q^2)a^k(m^2)+M^{OPE}
+O_2 M_n^{O2(1)}(Q^2)a(m^2)\,,\eqno(18)$$
$$M^{OPE}=O_2 M_n^{O2}(Q^2)+\sum\limits_{j=1}^2 O_3^j M_n^{O3,j}(Q^2)+
\sum\limits_{j=1}^7 O_4^j M_n^{O4,j}(Q^2)\,.$$
The explicit formula for $M_n^{(0)}$ can be obtained analytically:
$$M_n^{(0)}={1\over{(4m^2)^n}}{3\sqrt{\pi}(n-1)!\over
{4\Gamma(n+{5\over 2})}}\,{}_2F_1(n,1+n,5/2+n,-y)\,,$$
where $y=Q^2/(4m^2)$.
The next perturbative terms in (18) we calculate numerically (see (3)):
$$M_n^{(k)}(Q^2)={1\over{(4m^2)^n}}\int\limits_1^\infty{R^{(k)}(r,m^2)\,dr
\over{(r+y)^{n+1}}}\,.$$
Here $R^{(1)}$ is given by (6), for the  components of $R^{(2)}$ see (7),
(8),(9) and (15),(10),(11),(14).
The operator product expansion moments can be easily obtained from (16):
$$M_n^{O2}={1\over{(4m^2)^{n+2}}}\Big({1\over{3y^{n+1}}}-
{(3+n)!\,\Gamma(7/2)\over{15\,\Gamma(7/2+n)}}\,
{}_2F_1(1+n,4+n,7/2+n,-y)\Big)\,,$$
$$M_n^{O3,j}={4\over{27(4m^2)^{n+3}}}
\Big({c_{3,0}^j\over{y^{n+1}}}+\sum\limits_{i=1}^5
c_{3,i}^j {(1+i+n)!\,\Gamma(9/2)\over{(1+i)!\,\Gamma(9/2+n)}}\,
{}_2F_1(1+n,2+i+n,9/2+n,-y)\Big)\,,$$
$$M_n^{O4,j}={4\over{27(4m^2)^{n+4}}}
\Big({c_{4,0}^j\over{y^{n+1}}}+\sum\limits_{i=1}^7
c_{4,i}^j {(2+i+n)!\,\Gamma(11/2)\over{(2+i)!\,\Gamma(11/2+n)}}\,
{}_2F_1(1+n,3+i+n,11/2+n,-y)\Big)\,.$$
\indent As for $\alpha_s$-correction to the operator $O_2$,
the corresponding moments can be obtained by numerical differentiation of 
(17):
$$M_n^{O2(1)}={1\over{(4m^2)^2\,n!}}(-{d\over{dQ^2}})^n{P^A(y)\over
{8y^2(1+y)}}\,,$$
$y=Q^2/(4m^2)$.\\
\indent Calculating the numerical values of the moments one can easily
find that the $\alpha_s$-corrections to the moments are very large
and in fact the series (18) is divergent. The traditional solution of this
problem is the mass redefinition.  
Indeed, the pole mass in the above formulas has the meaning
of the mass of free quark and is ill defined quantity in the
charmonium sum rules.  This problem was discussed in details in
\cite{iof-zya}. Following \cite{iof-zya}, we consider
$\overline{\rm MS}$ scheme. The corresponding mass
${\bar m}$ is taken at the scale $\mu^2={\bar m}^2$:
${\bar m}={\bar m}({\bar m}^2)$. \\
\indent The pole mass $m$ can be expressed in terms of ${\bar m}$ as the
perturbative series:
$$m^2={\bar m}^2\Big(1+\sum\limits_{k=1,2,...}K_n a^k({\bar m})\Big)\,,$$
where $K_1,\,K_2,\,K_3$ were obtained in \cite{kkk}:
$$K_1=8/3\,,\;\;\;K_2=22.4162\,,\;\;\;K_3=260.526\,.$$
(These numbers are given for $n_l=3$.) \\
\indent Now we reexpand the series (18) in terms of the mass $\bar{m}$:
$$M_n(Q^2)=\sum\limits_{k=0}^2 {\bar M}_n^{(k)}(Q^2)a^k(m^2)+{\bar M}^{OPE}
+O_2 {\bar M}_n^{O2(1)}(Q^2)a(m^2)\,.\eqno(19)$$
Here
$$\displaylines{{\bar M}_n^{(0)}(Q^2)=M_n^{(0)}\,,\hfill}$$
$$\displaylines{{\bar M}_n^{(1)}(Q^2)=M_n^{(1)}-K_1 n M_n^{(0)}
+K_1(n+1)\,Q^2M_{n+1}^{(0)}\,,\hfill}$$
$$\displaylines{{\bar M}_n^{(2)}(Q^2)=M_n^{(2)}-K_1 n M_n^{(1)}
+K_1(n+1)\,Q^2M_{n+1}^{(1)}+n \Big({K_1^2\over 2} (n+1) - K_2 \Big)
  M_n^{(0)}+\hfill}$$
$$\displaylines{\hfill
+(n+1) \Big( K_2-K_1^2(n+1)\Big) Q^2M_{n+1}^{(0)}+{K_1^2\over 2}
(n+1)(n+2) \,Q^4 M_{n+2}^{(0)}\,,}$$
$$\displaylines{{\bar M}_n^{O2}(Q^2)=M_n^{O2}\,,
{\bar M}_n^{Ok,j}(Q^2)=M_n^{Ok,j}\,,\;\;k=3,4,\;\;\;
{\bar M}_n^{OPE}(Q^2)=M_n^{OPE}\,,\hfill}$$
$$\displaylines{{\bar M}_n^{O2(1)}(Q^2)=M_n^{O2(1)}-K_1 (n+2) M_n^{O2} +
K_1(n+1)\,Q^2M_{n+1}^{O2}\,.\hfill}$$
All moments in the right hand sides of these equations are computed with
the $\overline{\rm MS}$ mass ${\bar m}$. \\
\indent At the some other scale $\mu^2$
the function ${\bar M}_n^{(2)}$ changes
$${\bar M}_n^{(2)}(Q^2)\to{\bar M}_n^{(2)}(Q^2) \,+
{\bar M}_n^{(1)}(Q^2)\, \beta_0 \ln{ \mu^2\over{\bar m}^2}\,,\;\;\;
a({\bar m}^2)\to a(\mu^2)\eqno(20)$$
to ensure the scale independence at order $\alpha_s^2$.
\section{Restrictions on the gluon condensate value}
\indent To analyse the obtained data we introduce the
dimensionless ratio $r_n$ of the moments (19):
$$r_n={M_{n-1}(Q^2)\over{4{\bar m}^2M_{n}(Q^2)}}=
{m_\chi^2+Q^2\over{4{\bar m}^2}}+\delta\,\eqno(21).$$
Here $\delta$ stands for the continuum contribution. 
At large $n$ $\delta$ tends
to zero: $n\to\infty,\;\;\delta\to 0$. The theoretical ratio of the 
moments depends on
$Q^2$, quark mass ${\bar m}$, QCD coupling and condensates. First of all,
let us fix $Q^2$. At $Q^2=0$ the perturbative corrections as well as the
higher terms of the operator product expansion series are very large. 
On the other hand, at large $Q^2$, $Q^2/(4{\bar m}^2)\ge 3$, the
effective expansion parameter $a\beta_0\ln(Q^2/{\bar m}^2)$ in (20)
becomes large. In papers \cite{iof-zya},\cite{zya}
the values $Q^2/(4{\bar m}^2)=1$,
$Q^2/(4{\bar m}^2)=2$ were used. In the present paper we work at
$Q^2/(4{\bar m}^2)=2$. \\
\indent As for the QCD coupling constant, from the hadronic $\tau$-decay we
know \cite{gesh-iof-zya}:
$$\alpha_s(m_\tau^2)\,=\,0.33\pm 0.03\,,\eqno(22)$$
$m_\tau=1.777 \,{\rm GeV}$ is the mass of the $\tau$-lepton. To obtain
$\alpha_s$ at any other scale we solve numerically the renormalization
group equation. The choice of the scale is discussed  in \cite{iof-zya}:
$$\mu^2=Q^2+{\bar m}^2\,.\eqno(23)$$ \indent The value of the
$c$-quark mass ${\bar m}$ is determined in \cite{iof-zya},\cite{zya}
with high accuracy: $${\bar m}=1.275\pm 0.015\,{\rm GeV}$$ (this
value is taken from \cite{iof-zya}). Now theoretical $r_n$ depends on the
condensates only.\\ 
\indent Using the value
$m_\chi=3.51051\pm 0.00012\,{\rm GeV}$ \cite{pdg} we find the
phenomenological estimation of $r_n$ in (21):
$$r_n={m_\chi^2+Q^2\over{4{\bar m}^2}}+\delta=3.90\pm 0.05+\delta\,.
\eqno(24)$$ 
The uncertainty appears mainly due to the error in the quark mass value. \\
\indent Now let us look on the table of the moments.
$$\begin{array}{cccccc}
 n& a{\bar M}^{(1)}/{\bar M}^{(0)}& a^2{\bar M}^{(2)}/{\bar M}^{(0)}&
{\bar M}^{OPE}/{\bar M}^{(0)}& a{\bar M}^{O2(1)}/{\bar M}^{O2}\\
 4& 0.097& 0.032& 0.083& -0.83 \\
 5& 0.070& 0.018& 0.20& -0.51 \\
 6& 0.041& 0.0013& 0.44& -0.27 \\
 7& 0.0087& -0.017& 0.91& -0.065 \\
 8& -0.026& -0.037& 1.80& 0.12 \\
 9& -0.061& -0.058& 3.47& 0.30 \\
\end{array}
$$
In this table  $Q^2/(4{\bar m}^2)=2$,
$\langle{\alpha_s\over\pi}G^2\rangle/(4{\bar m}^2)=0.0002$ and scale (23)
are chosen. \\ Operators $O_3^k,\,O_4^k$ are excluded. \\
\indent The values in all the  columns in the certain line
should be small enough ($<0.5$ in magnitude), besides,
the values in the third column should be smaller than in the second one.
All these requirements ensure the convergence of the series (19). Thus we see
that there is very narrow range of $n$: $n=5-6$, and one can construct
the only ratio (21):
$$r_6={M_5\over{4{\bar m}^2M_{6}}}\,.$$
For $Q^2/(4{\bar m}^2)=1$ there are no appropriate $n$ at all. \\
\indent The values of the higher order operators, $O_3^k,\,O_4^k$, can be
estimated with the help of factorization hypothesis or instanton model
(for details see \cite{zya}).
In the instanton consideration
$$O^1_3={12\over 5\rho_c^2}\,O_2\,,~~~~ O^2_3=0 \,$$
$$\left( O_4^1, \ldots , O_4^7 \right) = \left( 4,8,3,4,0,8,0 \right)
{16\over 7\rho_c^4} \,O_2\,.$$
Here instanton radius $\rho_c=2.5\,{\rm GeV}^{-1}$. The instanton
concentration $n_0$ is connected with the gluon condensate:
$\langle{\alpha_s\over\pi}G^2\rangle=32n_0$. \\
\indent In the frame of the factorization hypothesis
$$\left( O_4^1, \ldots , O_4^7 \right) =
\left(\, {65\over 144}, \, {5\over 16}, \, {19\over 72}, \, {1\over 16},\,
 0,\, {1\over 8},\, 0\, \right) (O_2)^2\,.$$
\indent However, at $n=5-6$ both models give negligible corrections. \\
\indent At first we put $\delta=0$ in (24).
Then $3.85\le r_6 \le 3.95$. Now we find the
interval of the gluon condensate values, where these restrictions hold:
$$0.005\,{\rm GeV}^4 \le \langle{\alpha_s\over\pi}G^2\rangle \le
 0.006\,{\rm GeV}^4\,.$$
\indent The uncertainty in the coupling constant 
and variation of the scale
($\mu^2=Q^2$ or $\mu^2=Q^2+2{\bar m}^2$) give negligible correction to
 the value of gluon condensate. \\
\indent Now let us try to evaluate $\delta$. Unlike
to the vector channel, the mass of the second resonance with
$J^{PC}=1^{++}$ is unknown. If we suppose that the differences
between two lowest resonances in vector and axial-vector channels
are approximately equal (about $0.6\,{\rm GeV}$), the continuum
threshold can be evaluated: $s_0\approx 4.1\,{\rm GeV}$. Now one
can compare the integrals like (3) in the intervals $[4{\bar m}^2,\infty)$
and $[s_0^2,\infty)$ to evaluate the continuum
contribution to the certain moment. In such way we find:
$$\delta< 0.2\,.$$
It is especially important that $\delta$ is positive, $\delta>0$. \\
\indent It turns out that in the considered sum rules the
increase  of $r_6$ results in decrease of the value
$\langle{\alpha_s\over\pi}G^2\rangle$ (see fig.1). 
That is why introducing the positive
$\delta$ in (24) we obtain the smaller values of the gluon condensate. \\
\indent In some papers in the sum rules for heavy quarks the Coulomb-like
corrections  are  summed up. This is legitimate way for the 
nonrelativistic problems only. However, in our case the quark velocities
$v\approx\sqrt{(1+Q^2/(4{\bar m}^2))/n}\approx 0.7$ are large enough, and
the nonrelativistic corrections are not dominating. Detailed consideration 
of this question can be found in \cite{iof-zya}. \\
\indent Our final result is:
$$\langle{\alpha_s\over\pi}G^2\rangle=(0.005+0.001-0.004)\,{\rm
GeV}^4\,. \eqno(25)$$ 
Note again that the continuum (i.e.
higher states) does not affect the upper limit in (25), which can be
considered as quite reliable. As for the lower limit, it depends
on $\delta$, for which one has the estimation only. Therefore, even zero
value of the condensate can not be certainly excluded. \\ 
\indent
Our result (25) is in agreement with the values of the gluon
condensate, obtained in \cite{iof-zya}
($\langle{\alpha_s\over\pi}G^2\rangle=0.009\pm 0.007\,{\rm
GeV}^4$) and in \cite{zya}
($\langle{\alpha_s\over\pi}G^2\rangle<0.008\,{\rm GeV}^4$).
However, it is significantly smaller than $0.012\,{\rm
GeV}^4$, originally obtained in \cite{svz}.
\\ \\
\indent The author is thankful to B.L.\,Ioffe for the formulation of
the problem and fruitful discussions and to K.N.\,Zyablyuk for
helpful discussions. \\
\indent The work is supported in part by
grants INTAS 2000 Project 587 and RFFI 03-02-16209.

\begin{figure}[h]
\epsfxsize=10.0cm
\epsfbox{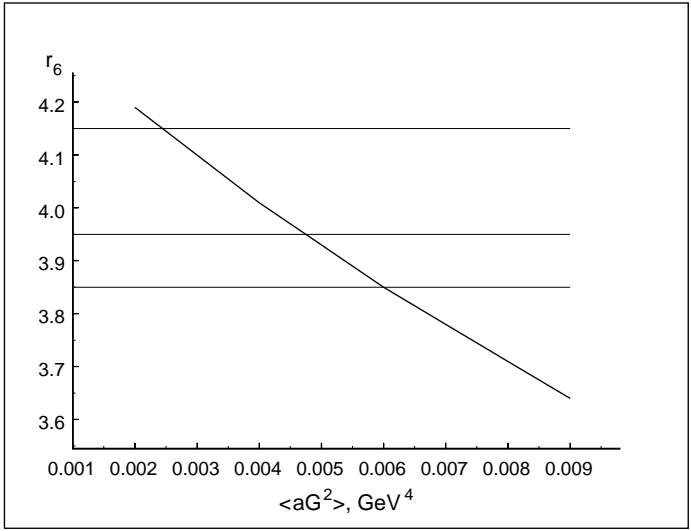}
\caption{\small{The ratio $r_6$ as the function of the gluon condensate. 
Horizontal lines denote the limits in (24).}}
\end{figure}

\end{document}